\documentclass[a4paper]{article}
\usepackage{RR}
\RRNo{6178}
\usepackage{hyperref}
\usepackage{amsmath,amssymb,graphicx,subfigure}
\usepackage{epsfig}
\usepackage{epic}
\usepackage{eepic}
\usepackage{graphics}

\usepackage{amscd}
\usepackage{latexsym}
\usepackage{tabularx}
\usepackage{a4wide}
\usepackage{here}
\usepackage{color}
\usepackage{enumerate}

\newtheorem{theorem}{Theorem}[section]
\newtheorem{corollary}[theorem]{Corollary}

\newtheorem{proposition}[theorem]{Proposition}
\newtheorem{lemma}[theorem]{Lemma}
\newtheorem{conjecture}[theorem]{Conjecture}

\newcommand{\ch}{\mathit{ch}}
\newcommand{\dist}{\mathit{dist}}

\newenvironment{proof}{\begin{trivlist}\item[]{\bf Proof}\mbox{ \ }}%
        {\qquad\hspace*{\fill}$\Box$\end{trivlist}}
\newcommand{\qite}[1]{\noindent\leavevmode\hangindent1.5\parindent%
        \noindent\hbox to1.5\parindent{#1\hss}\ignorespaces}

\RRdate{May 2007}
\RRauthor{
Omid Amini\thanks{Projet Mascotte, CNRS/INRIA/UNSA,
INRIA Sophia-Antipolis,
2004 route des Lucioles BP 93,
06902 Sophia-Antipolis Cedex,
France. {\tt oamini@sophia.inria.fr}}
  \and
Louis Esperet\thanks{LaBRI, Universit\'e Bordeaux 1, Talence, France. {\tt esperet@labri.fr}}
\and
Jan van den Heuvel \thanks{Centre for Discrete and Applicable Mathematics,
    Department of Mathematics, London School of Economics, London, U.K. {\tt jan@maths.lse.ac.uk}}
}
\authorhead{Amini \& Esperet \& v.d. Heuvel }
\RRtitle{Coloration frugale des graphes}
\RRetitle{Frugal Colouring of Graphs}
\titlehead{Frugal Colouring of Graphs}
\RRnote{The research for this paper was done during a visit of
  LE and JvdH to the Mascotte research group at INRIA Sophia-Antipolis. The
  authors like to thank the members of Mascotte for their hospitality.
   
 \hskip7.5pt JvdH's visit was partly supported by a grant from the British
  Council.}
\RRresume{Une coloration $k$-frugale d'un graphe $G$ est une coloration propre des sommets de $G$ telle que chaque couleur appara\^it au plus~$k$ fois 
au voisinage de chaque sommet. Ce type de coloration a \'et\'e introduit pour la premi\`ere fois par Hind. Moloy et Reed en 1997. Dans cet article, on \'etudie le nombre chromatique frugal des graphes planaires en g\'en\'eral, ceux de maille assez grande, et les graphes planaires ext\'erieurs. On relie ce param\`etre \`a plusieurs autres param\`etres bien connus comme la coloration du carr\'e, la coloration cyclique, et les $L(p,q)-$\'etiquetages des graphes planaires. On \'etudie aussi la version ar\^ete coloration frugale des multigraphes. 
}
\RRabstract{ A $k$-frugal colouring of a graph~$G$ is a proper colouring of the vertices
of~$G$ such that no colour appears more than~$k$ times in the neighbourhood
of a vertex. This type of colouring was introduced by Hind, Molloy and Reed
in 1997. In this paper, we study the frugal chromatic number of planar
graphs, planar graphs with large girth, and outerplanar graphs, and relate
this parameter with several well-studied colourings, such as colouring of
the square, cyclic colouring, and $L(p,q)$-labelling. We also study frugal
edge-colourings of multigraphs.
}
\RRmotcle{allocation de fr\'equences, coloration de graphes}
\RRkeyword{wavelength assignment, graph colouring}
\RRprojets{MASCOTTE}
\RRtheme{\THCom} 
\URSophia 
\begin{document}
\makeRR   

\section{Introduction}

Most of the terminology and notation we use in this paper is standard and
can be found in any text book on graph theory (\,such as~\cite{BoMu76}
or~\cite{Die05}\,). All our graphs and multigraphs will be finite. A
\emph{multigraph} can have multiple edges; a \emph{graph} is supposed to be
simple; loops are not allowed.

For an integer $k\ge1$, a \emph{$k$-frugal colouring} of a graph~$G$ is a
proper vertex colouring of~$G$ (\,i.e., adjacent vertices get a different
colour\,) such that no colour appears more than~$k$ times in the
neighbourhood of any vertex. The least number of colours in a $k$-frugal
colouring of $G$ is called the \emph{$k$-frugal chromatic number},
denoted~$\chi_k(G)$. Clearly, $\chi_1(G)$ is the chromatic number of the
square of~$G$; and for~$k$ at least the maximum degree of~$G$, $\chi_k(G)$
is the usual chromatic number of~$G$.

A \emph{$k$-frugal edge colouring} of a multigraph~$G$ is a (\,possibly
improper\,) colouring of the edges of~$G$ such that no colour appears more
than~$k$ times on the edges incident with a vertex. The least number of
colours in a $k$-frugal edge colouring of~$G$, the \emph{$k$-frugal edge
  chromatic number} (\,or \emph{$k$-frugal chromatic index}\,), is denoted
by~$\chi'_k(G)$. Remark that for $k=1$ we have $\chi'_1(G)=\chi'(G)$, the
normal chromatic index of~$G$.

When considering the possibility that each vertex or edge has a list of
available colours, we enter the area of \emph{frugal list (\,edge\,)
  colourings}.

Frugal vertex colourings were introduced by Hind \emph{et
  al}~\cite{HMR97,HMR99}, as a tool towards improving results about the
\emph{total chromatic number} of a graph. One of their results is that a
graph with large enough maximum degree~$\Delta$ has a
$(\log^8\!\Delta)$-frugal colouring using at most $\Delta+1$ colours. They
also show that there exist graphs for which a
$\bigl(\frac{\log\Delta}{\log\log\Delta}\bigr)$-frugal colouring cannot be
achieved using only $O(\Delta)$ colours.

Our aim in this note is to study some aspects of frugal colourings and
frugal list colourings in their own right. In the first part we consider
frugal vertex colourings of planar graphs. We show that for planar graphs,
frugal colouring are closely related to several other aspects that have
been the topic of extensive research the last couple of years. In
particular, we exhibit close connections with colouring the square, cyclic
colourings, and $L(p,q)$-labellings.

In the final section we derive some results on frugal edge colourings of
multigraphs in general.

\subsection{Further notation and definitions}

Given a graph~$G$, the \emph{square of~$G$}, denoted~$G^2$, is the graph
with the same vertex set as~$G$ and with an edge between any two different
vertices that have distance at most two in~$G$. We always assume that
colours are integers, which allows us to talk about the ``distance''
$|\gamma_1-\gamma_2|$ of two colours~$\gamma_1,\gamma_2$.

The \emph{chromatic number} of~$G$, denoted~$\chi(G)$, is the minimum
number of colours required so that we can properly colour its vertices
using those colours. A \emph{$t$-list assignment~$L$ on the vertices} of a
graph is a function which assigns to each vertex~$v$ of the multigraph a
list~$L(v)$ of~$t$ prescribed integers. The \emph{list chromatic number} or
\emph{choice number}~$\mathit{ch}(G)$ is the minimum value~$t$, so that for
each $t$-list assignment on the vertices, we can find a proper colouring in
which each vertex gets assigned a colour from its own private list.

We introduced \emph{$k$-frugal colouring} and the \emph{$k$-frugal
  chromatic number}~$\chi_k(G)$ in the introductory part. In a similar way
we can define \emph{$k$-frugal list colouring} and the \emph{$k$-frugal
  choice number}~$\ch_k(G)$.

Further definitions on edge colourings will appear in the final section.

\section{Frugal Colouring of Planar Graphs}\label{sec-planar}

In the next four sections we consider $k$-frugal (\,list\,) colourings of
planar graphs. For a large part, our work in that area is inspired by a
well-known conjecture of Wegner on the chromatic number of squares of
planar graphs. If~$G$ has maximum degree~$\Delta$, then a vertex colouring
of its square will need at least $\Delta+1$ colours, but the greedy
algorithm shows it is always possible with $\Delta^2+1$ colours. Diameter
two cages such as the 5-cycle, the Petersen graph and the Hoffman-Singleton
graph (\,see \cite[page~239]{BoMu76}\,) show that there exist graphs that
in fact require $\Delta^2+1$ colours.

For planar graphs, Wegner conjectured that far less than $\Delta^2+1$
colours should suffice.

\begin{conjecture}[\,Wegner~\cite{Weg77}\,]\mbox{}\\
  For a planar graph~$G$ of maximum degree $\Delta(G)\ge8$ we have
  $\chi(G^2)\le\bigl\lfloor\frac32\,\Delta(G)\bigr\rfloor+1$.
\end{conjecture}

\begin{figure}[!hbtp]
  \begin{center}
    \unitlength0.52mm
    \begin{picture}(82,94)(-50,-45)
      \put(-50,0){\circle*{3}}\put(-44,0){\circle*{3}}
      \put(-38,0){\circle*{3}}\put(-33,0){\circle*{1}}
      \put(-29,0){\circle*{1}}\put(-25,0){\circle*{1}}
      \put(-21,0){\circle*{1}}\put(-16,0){\circle*{3}}
      \put(25,43.30){\circle*{3}}\put(22,38.11){\circle*{3}}
      \put(19,32.91){\circle*{3}}\put(16.5,28.58){\circle*{1}}
      \put(14.5,25.11){\circle*{1}}\put(12.5,21.65){\circle*{1}}
      \put(10.5,18.19){\circle*{1}}\put(8,13.86){\circle*{3}}
      \put(25,-43.30){\circle*{3}}\put(22,-38.11){\circle*{3}}
      \put(19,-32.91){\circle*{3}}\put(16.5,-28.58){\circle*{1}}
      \put(14.5,-25.11){\circle*{1}}\put(12.5,-21.65){\circle*{1}}
      \put(10.5,-18.19){\circle*{1}}\put(8,-13.86){\circle*{3}}
      \put(-16,27.71){\circle*{3.5}}\put(-16,-27.71){\circle*{3.5}}
      \put(32,0){\circle*{3.5}}
      \qbezier(-16,27.71)(-16,13.855)(-16,0)
      \qbezier(-16,27.71)(-27,13.855)(-38,0)
      \qbezier(-16,27.71)(-30,13.855)(-44,0)
      \qbezier(-16,27.71)(-33,13.855)(-50,0)
      \qbezier(-16,-27.71)(-16,-13.855)(-16,0)
      \qbezier(-16,-27.71)(-27,-13.855)(-38,0)
      \qbezier(-16,-27.71)(-30,-13.855)(-44,0)
      \qbezier(-16,-27.71)(-33,-13.855)(-50,0)
      \qbezier(-16,27.71)(-4,20.785)(8,13.86)
      \qbezier(-16,27.71)(1.5,29.95)(19,32.91)
      \qbezier(-16,27.71)(3,32.91)(22,38.11)
      \qbezier(-16,27.71)(4.5,35.505)(25,43.30)
      \qbezier(-16,-27.71)(-4,-20.785)(8,-13.86)
      \qbezier(-16,-27.71)(1.5,-29.95)(19,-32.91)
      \qbezier(-16,-27.71)(3,-32.91)(22,-38.11)
      \qbezier(-16,-27.71)(4.5,-35.505)(25,-43.30)
      \qbezier(32,0)(20,6.93)(8,13.86)
      \qbezier(32,0)(25.5,16.455)(19,32.91)
      \qbezier(32,0)(27,19.055)(22,38.11)
      \qbezier(32,0)(28.5,21.65)(25,43.30)
      \qbezier(32,0)(20,-6.93)(8,-13.86)
      \qbezier(32,0)(25.5,-16.455)(19,-32.91)
      \qbezier(32,0)(27,-19.055)(22,-38.11)
      \qbezier(32,0)(28.5,-21.65)(25,-43.30)
      \qbezier(-16,27.71)(0,0)(-16,-27.71)
      \put(-54,3.5){\makebox(0,0)[r]{$m-1$}}
      \put(-54,-3.5){\makebox(0,0)[r]{vertices}}
      \put(29,43.40){\makebox(0,0)[l]{$m$ vertices}}
      \put(29,-43.40){\makebox(0,0)[l]{$m$ vertices}}
      \put(38,0){\makebox(0,0){$z$}}
      \put(-19,32.91){\makebox(0,0){$x$}}
      \put(-19,-32.91){\makebox(0,0){$y$}}
    \end{picture}
    \caption{The planar graphs $G_m$.}\label{tightfig}
  \end{center}
\end{figure}
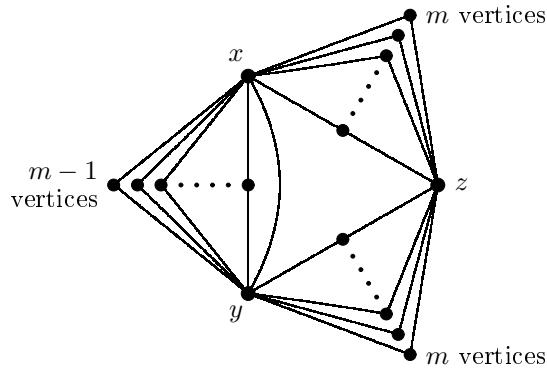

\noindent
Wegner also conjectured maximum values for the chromatic number of the
square of planar graph with maximum degree less than eight and gave
examples showing his bounds would be tight. For even $\Delta\ge8$, these
examples are sketched in Figure~\ref{tightfig}.

Inspired by Wegner's Conjecture, we conjecture the following bounds for the
$k$-frugal chromatic number of planar graphs.

\pagebreak[3]
\begin{conjecture}\label{conj1}\mbox{}\\*
  For any integer $k\ge1$ and planar graph $G$ with maximum degree
  $\Delta(G)\ge\max\,\{\,2\,k,\,8\,\}$ we have
  $$\chi_k(G)\:\le\:\left\{\begin{array}{ll}%
	\bigl\lfloor\frac{\Delta(G)-1}k\bigr\rfloor+3,&
		\text{if $k$ is even;}\\[1.5ex]
	\bigl\lfloor\frac{3\,\Delta(G)-2}{3\,k-1}\bigr\rfloor+3,&
		\text{if $k$ is odd.}\end{array}\right.$$
\end{conjecture}

\noindent
Note that the graphs~$G_m$ in Figure~\ref{tightfig} also show that the
bounds in this conjecture are best possible. The graph~$G_m$ has maximum
degree~$2\,m$. First consider a $k$-frugal colouring with $k=2\,\ell$ even.
We can use the same colour at most $\frac32\,k$ times on the vertices
of~$G_m$, and every colour that appears exactly $\frac32\,k=2\,\ell$ times
must appear exactly~$\ell$ times on each of the three sets of common
neighbours of~$x$ and~$y$, of~$x$ and~$z$, and of~$y$ and~$z$. So we can
take at most $\frac1\ell\,(m-1)=\frac1k\,(\Delta(G_m)-1)$ colours that are
used $\frac32\,k$ times. The graph that remains can be coloured using just
three colours.

If $k=2\,\ell+1$ is odd, then each colour can appear at most
$3\,\ell+1=\frac12\,(3\,k-1)$ times, and the only way to use a colour so
many times is by using it on the vertices in $V(G_m)\setminus\{x,y,z\}$.
Doing this at most
$\frac{3\,m-1}{(3\,k-1)/2}=\frac{3\,\Delta(G)-2}{3\,k-1}$ times, we are
left with a graph that can be coloured using three colours.

We next derive some upper bounds on the $k$-frugal chromatic number of
planar graphs. The first one is a simple extension of the approach
from~\cite{Van03}. In that paper, the following structural lemma is
derived.

\clearpage
\begin{lemma}[\,Van den Heuvel \& McGuinness
  \cite{Van03}\,]\label{vanheu}\mbox{}\\*
  Let~$G$ be a planar simple graph. Then there exists a vertex~$v$ with~$m$
  neighbours $v_1,\ldots,v_m$ with $d(v_1)\le\cdots\le d(v_m)$ such that
  one of the following holds\,:

  \smallskip
  \qite{(i)}$m\le2$;

  \smallskip
  \qite{(ii)}$m=3$ with $d(v_1)\le11$;

  \smallskip
  \qite{(iii)}$m=4$ with $d(v_1)\le7$ and $d(v_2)\le11$;

  \smallskip
  \qite{(iv)}$m=5$ with $d(v_1)\le6$, $d(v_2)\le7$, and $d(v_3)\le11$.
\end{lemma}

\noindent
Van den Heuvel and McGuinness~\cite{Van03} use this structural lemma to
prove that the chromatic number of the square of a planar graph is at most
$2\,\Delta+25$. Making some slight changes in their proof, it is not
difficult to obtain a first bound on~$\ch_k$ (\,and hence on~$\chi_k$\,)
for planar graphs.

\begin{theorem}\label{prop1}\mbox{}\\*
  For any planar graph~$G$ with $\Delta(G)\ge12$ and integer $k\ge1$ we
  have $\ch_k(G)\le\bigl\lfloor\frac{2\,\Delta(G)+19}k\bigr\rfloor+6$.
\end{theorem}

\begin{proof}We will prove that if a planar graph satisfies
  $\Delta(G)\le C$ for some $C\ge12$, then
  $\ch_k(G)\le\bigl\lfloor\frac{2\,C+19}k\bigr\rfloor+6$. We use induction
  on the number of vertices, noting that the result is obvious for small
  graphs. So let~$G$ be a graph with $|V(G)|>1$, choose $C\ge12$ so that
  $\Delta(G)\le C$, and assume each vertex~$v$ has a list~$L(v)$ of
  $\bigl\lfloor\frac{2\,C+19}k\bigr\rfloor+6$ colours. Take
  $v,v_1,\ldots,v_m$ as in Lemma~\ref{vanheu}. Contracting the edge~$vv_1$
  to a new vertex~$v'$ will result in a planar graph~$G'$ in which all
  vertices except~$v'$ have degree at most as much as they had in~$G$,
  while~$v'$ has degree at most~$\Delta(G)$ (\,for case~(i)\,) or at
  most~12. (\,for the cases (ii)\,--\,(iv)\,). In particular we have that
  $\Delta(G')\le C$. If we give~$v'$ the same list of colours as~$v_1$ had
  (\,all vertices in $V(G)\setminus\{v,v_1\}$ keep their list\,), then,
  using induction, $G'$ has a $k$-frugal colouring. Using the same
  colouring for~$G$, where~$v_1$ gets the colour~$v'$ had in~$G'$, we
  obtain a $k$-frugal colouring of~$G$ with the one deficit that~$v$ has no
  colour yet. But the number of colours forbidden for~$v$ are the colours
  on its neighbours, and for each neighbour~$v_i$, the colours that already
  appear~$k$ times around~$v_i$. So the number of forbidden colours is at
  most $m+\sum\limits_{i=1}^m\bigl\lfloor\frac{d(v_i)-1}k\bigr\rfloor$.
  Using the knowledge from the cases (i)\,--\,(iv), we get that
  $|L(v)|=\bigl\lfloor\frac{2\,C+19}k\bigr\rfloor+6$ is at least one more
  than this number of forbidden colours, hence we always can find an
  allowed colour for~$v$.
\end{proof}

\noindent
In the next section we will obtain (\,asymptotically\,) better results
based on more recent work on special labellings of planar graphs.



\section{Frugal Colouring and $L(p,q)$-Labelling}

Let $\dist(u,v)$ denote the distance between two vertices~$u,v$ in a graph.
For integers $p,q\ge0$, an \emph{$L(p,q)$-labelling of~$G$} is an
assignment~$f$ of integers to the vertices of~$G$ such that\,:

\smallskip
\qite{\quad$\bullet$}$|f(u)-f(v)|\ge p$, if $\dist(u,v)=1$, and

\smallskip
\qite{\quad$\bullet$}$|f(u)-f(v)|\ge q$, if $\dist(u,v)=2$.

\smallskip\noindent
The \emph{$\lambda_{p,q}$-number} of~$G$, denoted $\lambda_{p,q}(G)$, is
the smallest~$t$ such that there exists an $L(p,q)$-labellings of~$G$ using
labels from $1,2,\ldots,t$.\footnote{ The definition of $\lambda_{p,q}(G)$
  is not uniform across the literature. Many authors define it as the
  minimum distance between the largest and smallest label used, which gives
  a $\lambda$-value one less than with our definition. We chose our
  definitions since it means that $\lambda_{1,1}(G)=\chi(G^2)$, and since
  it fits more natural with the notion of list $L(p,q)$-labellings.}. Of
course we can also consider the list version of $L(p,q)$-labellings. Given
a graph~$G$, the \emph{list $\lambda_{p,q}$-number}, denoted
$\lambda^l_{p,q}(G)$, is the smallest integer~$t$ such that, for every
$t$-list assignment~$L$ on the vertices of~$G$, there exists an
$L(p,q)$-labelling~$f$ such that $f(v)\in L(v)$ for every vertex~$v$.

The following is an easy relation between frugal colourings and
$L(p,q)$-labellings.

\begin{proposition}\label{fr-pq}\mbox{}\\*
  For any graph~$G$ and integer $k\ge1$ we have
  $\chi_k(G)\le\bigl\lceil\frac1k\,\lambda_{k,1}(G)\bigr\rceil$ and
  $\ch_k(G)\le\bigl\lceil\frac1k\,\lambda^l_{k,1}(G)\bigr\rceil$.
\end{proposition}

\begin{proof}
  We only prove the second part, the first one can be done in a similar
  way. Set $\ell=\bigl\lceil\frac1k\,\lambda^l_{k,1}(G)\bigr\rceil$, and
  let~$L$ be an $\ell$-list assignment on the vertices of~$G$. Using that
  all elements in the lists are integers, we can define a new list
  assignment~$L^*$ by setting $L^*(v)=\bigcup_{x\in L(v)}\{k\,x,\,k\,x+1,$
  $\ldots,\,k\,x+k-1\}$. Then~$L^*$ is a $(k\,\ell)$-list assignment. Since
  $k\,\ell\ge\lambda^l_{k,1}(G)$, there exists an $L(k,1)$-labelling~$f^*$
  of~$G$ with $f^*(v)\in L^*(v)$ for all vertices~$v$. Define a new
  labelling~$f$ of~$G$ by taking
  $f(v)=\bigl\lfloor\frac1k\,f^*(v)\bigr\rfloor$. We immediately get that
  $f(v)\in L(v)$ for all~$v$. Since adjacent vertices received an
  $f^*$-label at least~$k$ apart, their $f$-labels are different. Also, all
  vertices in a neighbourhood of a vertex~$v$ received a different
  $f^*$-label. Since the map $x\mapsto\bigl\lfloor\frac1k\,x\bigr\rfloor$
  maps at most~$k$ different integers~$x$ to the same image, each $f$-label
  can appear at most~$k$ times in each neighbourhood. So~$f$ is a
  $k$-frugal colouring using labels from each vertex' list. This proves
  that $\ch_k(G)\le\ell$, as required.
\end{proof}

\noindent
We will combine this proposition with the following recent result.

\begin{theorem}[\,Havet \emph{et al} \cite{HHMR}\,]\label{th-weg}\mbox{}\\*
  For each $\epsilon>0$, there exists an integer~$\Delta_\epsilon$ so that
  the following holds. If~$G$ is a planar graph with maximum degree
  $\Delta(G)\ge\Delta_\epsilon$, and~$L$ is a list assignment so that each
  vertex gets a list of at least $(\frac32+\epsilon)\,\Delta(G)$ integers,
  then we can find a proper colouring of the square of~$G$ using colours
  from the lists. Moreover, we can take this proper colouring so that the
  colours on adjacent vertices of~$G$ differ by at least~$\Delta(G)^{1/4}$.
\end{theorem}

\noindent
In the terminology we introduced earlier, an immediate corollary is the
following.

\begin{corollary}\label{cor-1st}\mbox{}\\*
  Fix $\epsilon>0$ and an integer $k\ge1$. Then there exists an
  integer~$\Delta_\epsilon$ so that if~$G$ is a planar graph with maximum
  degree $\Delta(G)\ge\max\,\{\,\Delta_\epsilon,\,k^4\,\}$, then
  $\lambda^l_{k,1}(G)\le(\frac32+\epsilon)\,\Delta(G)$.
\end{corollary}

\noindent
Combining this with Proposition~\ref{fr-pq} gives the asymptotically best
upper bound for~$\chi_k$ and~$\ch_k$ for planar graphs we currently have.

\begin{corollary}\label{cor-2nd}\mbox{}\\*
  Fix $\epsilon>0$ and an integer $k\ge1$. Then there exists an
  integer~$\Delta_{\epsilon,k}$ so that if~$G$ is a planar graph with
  maximum degree $\Delta(G)\ge\Delta_{\epsilon,k}$, then
  $\ch_k(G)\le\frac{(3+\epsilon)\,\Delta(G)}{2\,k}$.
\end{corollary}

\noindent
In~\cite{MS05}, Molloy and Salavatipour proved that for any planar
graph~$G$, we have
$\lambda_{k,1}(G)\le\bigl\lceil\frac53\,\Delta(G)\bigr\rceil+18\,k+60$.
Together with Proposition~\ref{fr-pq}, this refines the result of
Proposition~\ref{prop1} and gives a better bound than
Corollary~\ref{cor-2nd} for small values of~$\Delta$. Note that this
corollary only concerns frugal colouring, and not frugal list colouring.

\begin{corollary}\label{cor-MS}\mbox{}\\*
  For any planar graph~$G$ and integer $k\ge1$, we have
  $\chi_k(G)\le\bigl\lceil\frac{5\,\Delta(G)+180}{3\,k}\bigr\rceil+18$.
\end{corollary}

\noindent
Proposition~\ref{fr-pq} has another corollary for planar graphs of
large girth that we describe below. The \emph{girth} of a graph is the
length of a shortest cycle in the graph.
 
In~\cite{LW03}, Lih and Wang proved that for planar graphs of large girth
the following holds\,:{

\smallskip
\qite{\quad$\bullet$}$\lambda_{p,q}(G)\le(2\,q-1)\,\Delta(G)+6\,p+12\,q-8$
for planar graphs of  girth at least six, and

\smallskip
\qite{\quad$\bullet$}$\lambda_{p,q}(G)\le(2\,q-1)\,\Delta(G)+6\,p+24\,q-14$
for planar graphs of girth at least five.

}\smallskip
Furthermore, Dvo\v{r}\'ak {\it et al}~\cite{DKNS} proved the following
tight bound for $(k,1)$-labellings of planar graphs of girth at least
seven, and of large degree.

\begin{theorem}[\,Dvo\v{r}\'ak \emph{et al} \cite{DKNS}\,]\mbox{}\\*
  Let~$G$ be a planar graph of girth at least seven, and maximum degree
  $\Delta(G)\ge190+2\,k$, for some integer $k\ge1$. Then we have
  $\lambda_{k,1}(G)\le\Delta(G)+2\,k-1$.

  Moreover, this bound is tight, i.e., there exist planar graphs which
  achieve the upper bound.
\end{theorem}

\noindent
A direct corollary of these results are the following bounds for planar
graphs with large girth.

\begin{corollary}\label{cor-girth}\mbox{}\\* 
  Let~$G$ be a planar graph with girth~$g$ and maximum degree~$\Delta(G)$.
  For any integer $k\ge1$, we have
  $$\chi_k(G)\:\le\:\left\{\begin{array}{rl}%
	\bigl\lceil\frac{\Delta(G)-1}k\bigr\rceil+2,&
		\text{if $g\ge7$ and $\Delta(G)\ge190+2\,k$;}\\[1ex]
	\bigl\lceil\frac{\Delta(G)+4}k\bigr\rceil+6,&
		\text{if $g\ge6$;}\\[1ex]
	\bigl\lceil\frac{\Delta(G)+10}k\bigr\rceil+6,&
		\text{if $g\ge5$.}\end{array}\right.$$
\end{corollary}

\section{Frugal Colouring of Outerplanar Graphs}

We now prove a variant of Conjecture \ref{conj1} for \emph{outerplanar
  graphs} (\,graphs that can be drawn in the plane so that all vertices are
lying on the outside face\,). For $k=1$, i.e., if we are colouring the
square of the graph, Hetherington and Woodall~\cite{HW06} proved the best
possible bound for outerplanar graphs~$G$\,: $\ch_1(G)\le\Delta(G)+2$ if
$\Delta(G)\ge3$, and $\ch_1(G)=\Delta(G)+1$ if $\Delta(G)\ge6$.

\begin{theorem}\label{outer1}\mbox{}\\*
  For any integer $k\ge2$ and any outerplanar graph~$G$ with maximum degree
  $\Delta(G)\ge3$, we have
  $\chi_k(G)\le\ch_k(G)\le\bigl\lfloor\frac{\Delta(G)-1}k\bigr\rfloor+3$.
\end{theorem}

\begin{proof}
  Esperet and Ochem~\cite{EO07} proved that any outerplanar graph contains
  a vertex~$u$ such that one of the following holds\,: (i)~$u$ has degree
  at most one; (ii)~$u$ has degree two and is adjacent to another vertex of
  degree two; or (iii)~$u$ has degree two and its neighbours~$v$ and~$w$
  are adjacent, and either~$v$ has degree three or~$v$ has degree four and
  its two other neighbours (\,i.e., distinct from~$u$ and~$w$\,) are
  adjacent.

  Let~$G$ be a counterexample to the theorem with minimum number of
  vertices, and let~$u$ be a vertex of~$G$ having one of the properties
  described above. By minimality of~$G$, there exists a $k$-frugal list
  colouring~$c$ of $G-u$ if the lists~$L(v)$ contain at least
  $\bigl\lfloor\frac{\Delta(G)-1}k\bigr\rfloor+3$ colours. If $u$ has
  property~(i) or~(ii), let~$t$ be the neighbour of $u$ whose degree is not
  necessarily bounded by two. It is easy to see that at most
  $2+\bigl\lfloor\frac{\Delta(G)-1}k\bigr\rfloor$ colours are forbidden
  for~$u$\,: the colours of the neighbours of~$u$ and the colours
  appearing~$k$ times in the neighbourhood of~$t$. If~$u$ has
  property~(iii), at most
  $2+\bigl\lfloor\frac{\Delta(G)-2}k\bigr\rfloor$ colours are forbidden
  for~$u$\,: the colours of the neighbours of~$u$ and the colours
  appearing~$k$ times in the neighbourhood of~$w$. Note that if~$v$ has
  degree four, its two other neighbours are adjacent and the $k$-frugality
  of~$v$ is respected since $k\ge2$. In all cases we found that at most
  $\bigl\lfloor\frac{\Delta(G)-1}k\bigr\rfloor+2$ colours are forbidden
  for~$u$. If~$u$ has a list with one more colour, we can extend~$c$ to a
  $k$-frugal list colouring of~$G$, contradicting the choice of~$G$.
\end{proof}

\noindent
We can refine this result in the case of 2-connected outerplanar graphs,
provided that $\Delta$ is large enough.

\begin{theorem}\label{outer2}
  For any integer $k\ge1$ and any 2-connected outerplanar graph~$G$ with
  maximum degree $\Delta(G)\ge7$, we have
  $\ch_k(G)\le\bigl\lfloor\frac{\Delta(G)-2}k\bigr\rfloor+3$.
\end{theorem}

\begin{proof}
  In Lih and Wang~\cite{LW06} it is proved that any 2-connected outerplanar
  graphs with maximum degree $\Delta\ge7$ contains a vertex~$u$ of degree
  two that has at most $\Delta-2$ vertices at distance exactly two.

  Let~$G$ be a counterexample to the theorem with minimum number of
  vertices, and let~$u$ be a vertex of~$G$ having the property described
  above, and let~$v$ and~$w$ be its neighbours. Let~$H$ be $G-u$ if the
  edge~$vw$ exists, or $G-u+vw$ otherwise. By minimality of~$G$, there is a
  $k$-frugal list colouring~$c$ of~$H$ if all lists contain at least
  $\bigl\lfloor\frac{\Delta(G)-2}k\bigr\rfloor+3$ colours. At most
  $\bigl\lfloor\frac{\Delta(G)-2}k\bigr\rfloor+2$ colours are forbidden
  for~$u$\,: the colours of~$v$ and~$w$, and the colours appearing~$k$
  times in their neighbourhood. So, the colouring~$c$ of $H$ can be
  extended to a $k$-frugal list colouring of~$G$, contradicting the choice
  of~$G$.
\end{proof}

\section{Frugal Colouring and Cyclic Colouring}

In this section, we discuss the link between frugal colouring and cyclic
colouring of plane graphs. A \emph{plane graph}~$G$ is a planar graph with
a prescribed planar embedding. The size (\,number of vertices in its
boundary\,) of a largest face of~$G$ is denoted by $\Delta^*(G)$.

A \emph{cyclic colouring} of a plane graph~$G$ is a vertex colouring of~$G$
such that any two vertices incident to the same face have distinct colours.
This concept was introduced by Ore and Plummer~\cite{OP69}, who also proved
that a plane graph has a cyclic colouring using at most $2\,\Delta^*$
colours. Borodin~\cite{Bor84} (\,see also Jensen and
Toft~\cite[page~37]{JeTo95}\,) conjectured that any plane graph has a
cyclic colouring with $\bigl\lfloor\frac32\,\Delta^*\bigr\rfloor$ colours,
and proved this conjecture for $\Delta^*=4$. The best known upper bound in
the general case is due to Sanders and Zhao~\cite{San01}, who proved that
any plane graph has a cyclic colouring with
$\bigl\lceil\frac53\,\Delta^*\bigr\rceil$ colours.

There appears to be a strong connection between bounds on colouring the
square of planar graphs and cyclic colourings of plane graphs. One should
only compare Wegner's conjecture in Section~\ref{sec-planar} with Borodin's
conjecture above, and the successive bounds obtained for each of these
connections. Nevertheless, the similar looking bounds for these types of
colourings have always required independent proofs. No explicit relation
that would make it possible to translate a result on one of the types of
colouring into a result for the other type, has ever been derived.

In this section we show that if there is an even $k\ge4$ so that Borodin's
conjecture holds for all plane graphs with $\Delta^*\le k$, and our
Conjecture~\ref{conj1} is true for the same value~$k$, then Wegner's
conjecture is true up to an additive constant factor.

\begin{theorem}\mbox{}\\*
  Let $k\ge4$ be an even integer such that every plane graph~$G$ with
  $\Delta^*(G)\le k$ has a cyclic colouring using at most $\frac32\,k$
  colours. Then, if~$G$ is a planar graph satisfying
  $\chi_k(G)\le\bigl\lfloor\frac{\Delta(G)-1}{k}\bigr\rfloor+3$, we
  also have
  $\chi(G^2)=\chi_1(G)\le\bigl\lfloor\frac32\,\Delta(G)\bigr\rfloor+
  \frac92\,k-1$.
\end{theorem}

\begin{proof}
  Let~$G$ be a planar graph with a given embedding and let $k\ge4$ be an
  even integer such that
  $t=\chi_k(G)\le\bigl\lfloor\frac{\Delta(G)-1}{k}\bigr\rfloor+3$. Consider
  an optimal $k$-frugal colouring~$c$ of~$G$, with colour classes
  $C_1,\ldots,C_t$. For $i=1,\ldots,t$, construct the graph~$G_i$ as
  follows\,: Firstly, $G_i$ has vertex set~$C_i$, which we assume to be
  embedded in the plane in the same way they were for~$G$. For each vertex
  $v\in V(G)\setminus C_i$ with exactly two neighbours in~$C_i$, we add an
  edge in~$G_i$ between these two neighbours. For a vertex $v\in
  V(G)\setminus C_i$ with $\ell\ge3$ neighbours in~$C_i$, let
  $x_1,\ldots,x_\ell$ be those neighbours in~$C_i$ in a cyclic order
  around~$v$ (\,determined by the plane embedding of~$G$\,). Now add edges
  $x_1x_2,x_2x_3,\ldots,x_{\ell-1}x_\ell$ and~$x_\ell x_1$ to~$G_i$. These
  edges will form a face of size~$\ell$ in the graph we have constructed so
  far. Call such a face a \emph{special face}. Note that since~$C_i$ is a
  colour class in a $k$-frugal colouring, this face has size at most~$k$.

  Do the above for all vertices $v\in V(G)\setminus C_i$ that have at least
  two neighbours in~$C_i$. The resulting graph is a plane graph with some
  faces labelled special. Add edges to triangulate all faces that are not
  special. The resulting graph is a plane graph with vertex set~$G_i$ and
  every face size at most~$k$. From the first hypothesis it follows that we
  can cyclicly colour each~$G_i$ with $\frac32\,k$ new colours. Since every
  two vertices in~$C_i$ that have a common neighbour in~$G$ are adjacent
  in~$G_i$ or are incident to the same (\,special\,) face, vertices
  in~$C_i$ that are adjacent in the square of~$G$ receive different
  colours. Hence, combining these~$t$ colourings, using different colours
  for each~$G_i$, we obtain a colouring of the square of~$G$, using at most
  $\frac32\,k\cdot\bigl(\bigl\lfloor\frac{\Delta(G)-1}{k}\bigr\rfloor+3\bigr)\le
	\bigl\lfloor\frac32\,\Delta\bigr\rfloor+\frac92\,k-1$ colours.
\end{proof}

\noindent
Since Borodin~\cite{Bor84} proved his cyclic colouring conjecture in the
case $\Delta^*=4$, we have the following corollary.

\begin{corollary}\mbox{}\\*
  If~$G$ is a planar graph so that
  $\chi_4(G)\le\bigl\lfloor\frac{\Delta(G)-1}{4}\bigr\rfloor+3$, then
  $\chi(G^2)\le\bigl\lfloor\frac32\,\Delta(G)\bigr\rfloor+17$.
\end{corollary}

\section{Frugal Edge Colouring}

An important element in the proof in~\cite{HHMR} of Theorem~\ref{th-weg}
mentioned earlier is the derivation of a relation between (\,list\,)
colouring square of planar graphs and edge (\,list\,) colourings of
multigraphs. Because of this, it seems to be opportune to have a short look
at a frugal variant of edge colourings of multigraphs in general.

If we need to properly colour the edges of a multigraph~$G$, the minimum
number of colours required is the \emph{chromatic index},
denoted~$\chi'(G)$. The \emph{list chromatic index~$\ch'(G)$} is defined
analogously as the minimum length of list that needs to be given to each
edge so that we can use colours from each edge's list to give a proper
colouring.

A \emph{$k$-frugal edge colouring} of a multigraph~$G$ is a (\,possibly
improper\,) colouring of the edges of~$G$ such that no colour appears more
than~$k$ times on the edges incident with a vertex. The least number of
colours in a $k$-frugal edge colouring of~$G$, the \emph{$k$-frugal edge
  chromatic number} (\,or \emph{$k$-frugal chromatic index}\,), is denoted
by~$\chi'_k(G)$.

Note that a $k$-frugal edge colouring of~$G$ is not the same as a
$k$-frugal colouring of the vertices of the line graph~$L(G)$ of~$G$. Since
the neighbourhood of any vertex in the line graph~$L(G)$ can be partitioned
into at most two cliques, every proper colouring of~$L(G)$ is also a
$k$-frugal colouring for $k\ge2$. A 1-frugal colouring of~$L(G)$ (\,i.e., a
vertex colouring of the square of~$L(G)$\,) would correspond to a proper
edge colouring of~$G$ in which each colour class induces a matching. Such
colourings are known as \emph{strong edge colourings}, see,
e.g.,~\cite{FF83}.

The list version of $k$-frugal edge colouring can also be defined in the
same way\,: given lists of size~$t$ for each edge of~$G$, one should be
able to find a $k$-frugal edge colouring such that the colour of each edge
belongs to its list. The smallest~$t$ with this property is called the
\emph{$k$-frugal edge choice number}, denoted~$\ch'_k(G)$.

Frugal edge colourings and its list version were studied under the name
\emph{improper edge-colourings} and \emph{improper L-edge-colourings} by
Hilton \emph{et al}~\cite{HSS01}.

It is obvious that the chromatic index and the edge choice numbers are
always at least the maximum degree~$\Delta$. The best possible upper bounds
in terms of the maximum degree only are given by the following results.

\begin{theorem}\label{edgecol-th}\mbox{}

  \qite{\rm(a)} For a simple graph~$G$ we have
  $\chi'(G)\le\Delta(G)+1$.\hfill {\rm\textbf{(\,Vizing \cite{Viz64}\,)}}

  \smallskip
  \qite{\rm(b)}For a multigraph~$G$ we have
  $\chi'(G)\le\bigl\lfloor\frac32\,\Delta(G)\bigr\rfloor$.\hfill
  {\rm\textbf{(\,Shannon \cite{Sh49}\,)}}

  \smallskip
  \qite{\rm(c)}For a bipartite multigraph~$G$ we have
  $\ch'(G)=\Delta(G)$.\hfill {\rm\textbf{(\,Galvin \cite{Gal95}\,)}}

  \smallskip
  \qite{\rm(d)}For a multigraph~$G$ we have
  $\ch'(G)\le\bigl\lfloor\frac32\,\Delta(G)\bigr\rfloor$.\hfill
  {\rm\textbf{(\,Borodin \emph{et al} \cite{BKW97}\,)}}
\end{theorem}

\noindent
We will use Theorem~\ref{edgecol-th}\,(c) and~(d) to prove two results on
the $k$-frugal chromatic index and the $k$-frugal choice number. The first
result shows that for even~$k$, the maximum degree completely determines
the values of these two numbers. This result was earlier proved
in~\cite{HSS01} in a slightly more general setting, involving a more
complicated proof.

\begin{theorem}[\,Hilton \emph{et al} \cite {HSS01}\,]\label{th-even}\mbox{}\\*
  Let~$G$ be a multigraph, and let~$k$ be an even integer. Then we have
  $\chi'_k(G)=\ch'_k(G)=\bigl\lceil\frac1k\,\Delta(G)\bigr\rceil$.
\end{theorem}

\begin{proof}
  It is obvious that
  $\ch'_k(G)\ge\chi'_k(G)\ge\bigl\lceil\frac1k\,\Delta\bigr\rceil$, so it
  suffices to prove $\ch'_k(G)\le\bigl\lceil\frac1k\,\Delta\bigr\rceil$.

  Let $k=2\,\ell$. Without loss of generality, we can assume~$\Delta$ is a
  multiple of~$k$ and~$G$ is a $\Delta$-regular multigraph. (\,Otherwise,
  we can add some new edges and, if necessary, some new vertices. If this
  larger multigraph is $k$-frugal edge choosable with lists of size
  $\bigl\lceil\frac1k\,\Delta\bigr\rceil$, then so is~$G$.\,) As~$k$, and
  hence~$\Delta$, is even, we can find an Euler tour in each component
  of~$G$. By given these tours a direction, we obtain an orientation~$D$ of
  the edges of~$G$ such that the in-degree and the out-degree of every
  vertex is $\frac12\,\Delta$. Let us define the bipartite multigraph
  $H=(V_1\cup V_2,E)$ as follows\,: $V_1,V_2$ are both copies of~$V(G)$.
  For every arc $(a,b)$ in~$D$, we add an edge between $a\in V_1$ and
  $b\in V_2$.

  Since~$D$ is a directed multigraph with in- and out-degree equal to
  $\frac12\,\Delta$, $H$ is a $(\frac12\,\Delta)$-regular bipartite
  multigraph. That means we can decompose the edges of~$H$ into
  $\frac12\,\Delta$ perfect matchings $M_1,M_2,\dots,M_{\Delta/2}$. Define
  disjoint subgraphs $H_1,H_2,\dots,H_\ell$ as follows\,: for
  $i=0,1,\ldots,\ell-1$ set $H_{i+1}=M_{\frac{i}k\Delta+1}\cup
  M_{\frac{i}k\Delta+2}\cdots\cup M_{\frac{i+1}k\Delta}$. Notice that
  each~$H_i$ is a bipartite multigraph of regular degree $\frac1k\,\Delta$.

  Now, suppose that each edge comes with a list of colours of size
  $\frac1k\,\Delta$. (\,If we had to add edges to make~$\Delta$ a multiple
  of~$k$ or the multigraph $\Delta$-regular, then give arbitrary lists to
  these edges.\,) Each subgraph~$H_i$ has maximum degree $\frac1k\,\Delta$,
  so by Galvin's theorem we can find a proper edge colouring of each~$H_i$
  such that the colour of each edge is inside its list. We claim that the
  same colouring of edges in~$G$ is $k$-frugal. For this we need the
  following observation\,:

  \begin{trivlist}
  \item[]\textbf{Observation} \ \emph{Let~$M$ be a matching in~$H$. Then
      the set of corresponding edges in~$G$ form a subgraph of maximum
      degree at most two.}
  \end{trivlist}
  To see this, remark that each vertex has two copies in~$H$\,: one
  in~$V_1$ and one in~$V_2$. The contribution of the edges of~$M$ to a
  vertex~$v$ in the original multigraph is then at most two, at most one
  from each copy of~$v$.

  To conclude, we observe that each colour class in~$H$ is the union of at
  most~$\ell$ matchings, one in each~$H_i$. So at each vertex, each colour
  class appears at most two times the number of~$H_i$'s, i.e., at most
  $2\,\ell=k$ times. This is exactly the $k$-frugality condition we set out
  to satisfy.
\end{proof}

\noindent
For odd values of~$k$ we give a tight upper bound of the $k$-frugal edge
chromatic number.

\begin{theorem}\label{th-odd}\mbox{}\\*
  Let~$k$ be an odd integer. Then we have
  $\bigl\lceil\frac{\Delta(G)}k\bigr\rceil\le\chi'_k(G)\le\ch'_k(G)\le
	\bigl\lceil\frac{3\,\Delta(G)}{3\,k-1}\bigr\rceil$.
\end{theorem}

\begin{proof}
  Again, all we have to prove is
  $\ch'_k(G)\le\bigl\lceil\frac{3\,\Delta(G)}{3\,k-1}\bigr\rceil$.

  Let $k=2\,\ell+1$. Since $3\,k-1$ is even and not divisible by three, we
  can again assume, without loss of generality, that~$\Delta$ is even and
  divisible by $3\,k-1$, and that~$G$ is $\Delta$-regular. Set
  $\Delta=m\,(3\,k-1)=6\,\ell\,m+2\,m$. Using the same idea as in the
  previous proof, we can decompose~$G$ into two subgraphs $G_1,G_2$,
  where~$G_1$ is $(6\,\ell\,m)$-regular and~$G_2$ is $(2\,m)$-regular.
  (\,Alternatively, we can use Petersen's Theorem~\cite{Pet91} that every
  even regular multigraph has a \mbox{2-factor}, to decompose the edge set
  in 2-factors, and combine these 2-factors appropriately.\,) Since
  $\frac1{2\,\ell}\cdot6\,\ell\,m=\frac3{3\,k-1}\,\Delta$, by
  Theorem~\ref{th-even} we know that~$G_1$ has a $2\,\ell$-frugal edge
  colouring using the colours from each edge's lists. Similarly we have
  $\frac32\cdot2\,m=\frac3{3\,k-1}\,\Delta$, and hence
  Theorem~\ref{edgecol-th}\,(d) guarantees that we can properly colour the
  edges of~$G_2$ using colours from those edges' lists. The combination of
  these two colourings is a $(2\,\ell+1)$-frugal list edge colouring, as
  required.
\end{proof}

\noindent
Note that Theorem~\ref{th-odd} is best possible\,: For $m\ge1$,
let~$T^{(m)}$ be the multigraph with three vertices and~$m$ parallel edges
between each pair. If $k=2\,\ell+1$ is odd, then the maximum number of
edges with the same colour a $k$-frugal edge colouring of~$T^{(m)}$ can
have is $3\,\ell+1$. Hence the minimum number of colours needed for a
$k$-frugal edge colouring is
$\Bigl\lceil\frac{3\,m}{3\,\ell+1}\Bigr\rceil=
	\Bigl\lceil\frac3{3\,k-1}\,\Delta(T^{(m)})\Bigr\rceil$.

\section{Discussion}

As this is one of the first papers on frugal colouring, many possible
directions for future research are still open. An intriguing question is
inspired by the results on frugal edge colouring in the previous section.
These results demonstrate an essential difference between even and odd~$k$
as far as $k$-frugal edge colouring is concerned. Based on what we think
are the extremal examples of planar graphs for $k$-frugal \emph{vertex}
colouring, also our Conjecture~\ref{conj1} gives different values for even
and odd~$k$. But for frugal vertex colourings of planar graphs in general
we have not been able to obtain results that are different for even and
odd~$k$. Most of our results for vertex colouring of planar graphs are
consequences of Proposition~\ref{fr-pq} and known results on
$L(k,1)$-labelling of planar graphs, for which no fundamental difference
between odd and even $k$ has ever been demonstrated. Hence, a major step
would be to prove that Proposition \ref{fr-pq} is far from tight when $k$
is even.

A second line of future research could be to investigate which classes of
graphs have \mbox{$k$-frugal} chromatic number equal to the minimum
possible value $\bigl\lceil\frac{\Delta}k\bigr\rceil+1$.
Corollary~\ref{cor-girth} and Theorems~\ref{outer1} and~\ref{outer2} give
bounds for planar graphs with large girth and outerplanar graphs with large
maximum degree that are very close to the best possible bound. We
conjecture that, in fact, planar graphs with large enough girth and
outerplanar graphs of large enough maximum degree do satisfy
$\chi_k(G)=\bigl\lceil\frac{\Delta(G)}k\bigr\rceil+1$ for all $k\ge1$.



\tableofcontents

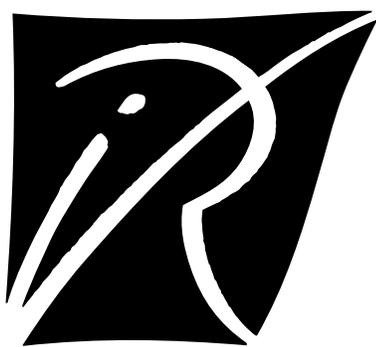
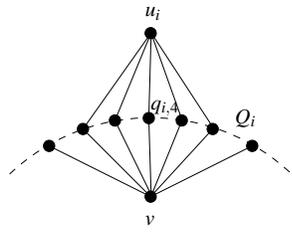
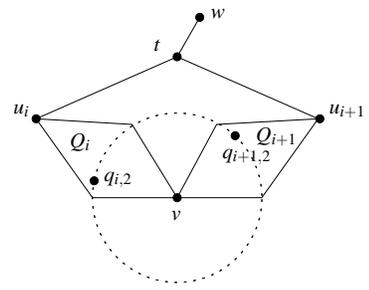
\end{document}